# 1,000 Days to First Light

## Construction of the Perth-Lowell Telescope Facility, 1968-71

Robert G. Hunt

MSc GDipEdSc GCertSpSc BAppScBiol

*rghunt@tpg.com.au*

## INTRODUCTION

The oldest known sky-watching construction is a 7,000-year-old rock pattern in a paddock in south-east Australia (Norris et al 2012). From this, to the European Southern Observatory's Overwhelmingly Large Telescope (ESOweb), and everything in between, optical, land-based observatories have but a few common features. Contemporary telescopes require a building site that is open, raised and away from large populations and such installations are worthwhile only when very specific science requirements are requested, justified, costed, approved and realised.

In 1966, one of only a handful of university/publicly funded Australian observatories (ASAweb), the Perth Observatory (PO) in Western Australia (WA), was relocated 25 km due east to the top of a nearby escarpment. The main research telescope at the new site is known as the 24-inch Perth-Lowell. After arriving onsite in 1971, this telescope played a significant role in planetary surveys (Baum 1974), the search for extra-terrestrial intelligence (SETI) (UWAweb1), exoplanet discovery (SNweb), supernovae searches (SNweb), meteorite astrometry (Burman 1992), planetary ring discoveries (Bowers 2017) and comet astrometrics (Jekabsons 1976). It is mounted under a motorised dome atop a tower and is a focal point for visitors who come from far and wide for historical and star-watching tours.

Construction of the Perth-Lowell facility has its origins at the National Aeronautics and Space Administration (NASA) and the Lowell Observatory (LO) in Flagstaff, Arizona. It was a high performer in planetary surveys of Mars, Jupiter and Saturn (PO Arch, Fisher 2016, Baum 1974, Baum et al. 1970). Mars was in opposition in 1969, 71 and 73 with the best viewing expected for the southern hemisphere in 1971, and an International Planetary Patrol Network (IPPN) of eight observatories was to supply important data for NASA.

The excitement of the Apollo decade of space travel began in 1961, and by 1969 NASA's IPPN was setting up shop. NASA engaged the LO who then wished to establish and operate the eight telescopes strategically placed around the globe (Baum et al 1970). The Perth-Lowell was very successful at collecting quality observations, especially helpful for the Viking Mars missions of the mid-1970s (see Vikweb).

The manufacturer of the telescope itself, Boller and Chivens (B&C), were a Pasadena, California company founded in 1946 (Scopeweb) and specialising in mechanical and optical equipment (B&Cweb). Negotiations between the PO, the LO, B&C, NASA, and several WA Government departments finally resulted in a very unique structure to house the telescope and as a result of an ongoing memorandum of understanding, the telescope is still at the PO in 2020.

A WA Heritage Council assessment describes the main PO building as a "…*well-resolved example of the Late Twentieth-Century International style*…" with "…*high social value*…" and "…*very high scientific value*…" …and with a "…*high degree of authenticity. Changes to the place have been limited to construction of the new telescope buildings throughout the grounds including the 24-inch Reflector Building in 1971*…" (HCWAweb).

The Perth-Lowell installation is a three-legged, concrete structure with a wrap-around external stair case and a 41 ft 7 inch (12 m) tall, concrete telescope pier that extends 10 ft into the ground and is exposed to the elements. For reasons discovered by the current research, this unusual design changed in height over several architectural iterations, and was variously influenced by predilection, science, budget, aesthetic licence and opportunism.

**AIM**

The radiant point for this investigation is an anecdote about the tower height, often repeated by Astronomy Officer Mr Greg Lowe to Star Viewing Night patrons of the PO. He "…*believe[d] that it was a result of the astronomers back in the late '60s taking advantage of the WA Government's largesse and lack of scientific know-how, in order to have the building made as high as possible to escape the shimmer at ground level*." (Lowe 2017). Further discussions about the uniqueness of the structure were uncommon and uncorroborated. Lowe is a retired, long serving, highly regarded employee of the PO and subsequent discussions with him and the Acting Government Astronomer at the time, Mr Ralph Martin, raised the question of how such subjective licence was given/taken to build the facility.

Why so tall? Avoiding ground thermals by raising the tower to the height of the forest canopy does not place it in lamina air flow and the cost of doing so far outweighs the cost of clearing some trees and placing it at a more modest height. Night- and day-tour hosts for the PO use the Perth-Lowell dome as one of their key talking points, not only for its astronomical achievements, but its physical housing, political story, instrumentation, and example as a unique piece of scientific hardware with real connections to economies and people.

The current work followed fabrication of the telescope by the Caltech spin-off business Boller & Chivens (B&Cweb), bankrolling by wealthy scientist/inventor Arnold Beckman (Galwas 2004), ocean freighting for six weeks, installation in a unique dome, and commissioning by NASA for the IPPN survey. It addressed the circumstances that lead to a local State Government architect deeming it "*pleasing*" (PO Arch) to raise the telescope to twice the specified height.

This research merged several timelines and synthesized the evidence trails. There are the individuals' trails: principal architect at the PWD, Mr Tadeusz Andrzejaczek (1915-1987), Director at the LO, Prof. William Alvin "Bill" Baum (1924-2012) and Acting Government Astronomer at the PO, Dr Bertrand "John" Harris (1925-1974). There was the telescope's trail: NASA, LO, B&C, transport, installation. And there were the local institutions' trails: Mt Stromlo Observatory, PO, WA Treasury and the PWD.

This research examined the 1,000 days prior to first light capture on April 8, 1971, of the Perth-Lowell telescope and seeks to show how WA's multi-cultural, international, science community produced the unique telescope installation. It also sought to show how the design of the tower was influenced by specific features of existing buildings around the world, as well as the personal and professional predilections of several key individuals.

**METHODS**

Inputs for the research came from:

- a personal, authorised face-to-face interview in the home of Andrzejaczek's son, Krysystof. The meeting was not digitally recorded but hand written notes of his responses may be regarded as a primary source. The contents of the interview critically informed understanding of why the well-connected Andrzejaczek had the imprimatur to make such decisions as he did;
- personal conversations with builders, surveyors, astronomers, researchers;
- personal emails to managers of observatories, history researchers, government departments, academics, scientists, engineers and artists;
- PO and LO archived documents;
- WA Government department records and archives;
- text books, PhDs, and high credibility web sites;
- NASA's abstract service, ADSABS (ADSABSweb); and
- an extensive internet search.

**RESULTS**

Fig.1 is a timeline of events for each of the two key personalities in this story. From 1915 to 1971, it shows how different their backgrounds were, but as will be seen, it hides the great similarities in their motivation, intelligence and professionalism. Andrzejaczek's life trail is on the left and Harris' is on the right. Some significant astronomical and historical dates are interspersed. This timeline should help the reader follow the initially very different, but then overlapping lives.

Tadeusz Andrzejaczek was born just after the end of WW I, in 1915 in Częstochowa, the second largest city in the southern Polish province of Silesia. Just a year before, Russia had relinquished, to the invading German Army, its 100-year rule over the region. Tens of thousands of inhabitants were forced to flee to the next province, and Andrzejaczek, having been born into very difficult family circumstances, was taken 400 km away by his grandparents to their home in the northern city of Bydgoszcz. Andrzejaczek completed his schooling there at around 18 years old before moving to the Polish capital, Warsaw. Whether or not he travelled around Poland during his teen years is unknown, but unlikely (Andrzejaczek 2017).

Concurrent with the post-war re-unification of Poland, and other geo-political re-shaping by the 1919 Treaty of Versailles, an architectural design movement was beginning to appear. The Modern Movement of the first half of the 20th century includes the International Style that originated in Europe around 1925. At the vanguard of this style were France (through Le Corbusier) and Germany (through Gropius and van der Rohe) (PHMCweb) and well-known examples include Le Corbusier's 1928 Villa Savoye in Paris and van der Rohe's 1929 Barcelona Pavillion in Spain. These were to become Andrzejaczek's 'world'.

Andrzejaczek's tumultuous young life was likely difficult, but he had a deep interest in 'form' and the built environment and, due to his love of all things French, would doubtless have been aware of the bourgeoning International Style in architecture (Andrzejaczek 2017). One of the observatories in his birth city of Częstochowa has some interesting resemblances to the Perth-Lowell telescope design. Spending the interbellum period in Warsaw, the talented, creative and adolescent Andrzejaczek joined the Polish Army in 1939 at the age of 24. Army life was not kind and while still in Warsaw and, within a week of signing up, he suffered a shrapnel injury to his knee. Polish captives were not treated strictly as prisoners of war by Germany, but he did spend the rest of the war in German concentration camps.

Bertrand John Harris was born 1925 in Patchem, Sussex in a tranquil, green south east corner of England. Harris' young life was quite different to Andrzejaczek's. Harris attended the Royal Grammar School of King Edward VI in Guildford. He played rugby, cricket and chess. He won a Magnus mathematics prize, took part in debating and graduated in 1941 (ADBHweb). By age 21 he had served 3 years in the Royal Navy and had become employed by the Royal Greenwich Observatory, London.

After an extraordinary tale of escape from the German holding camp system, Andrzejaczek found himself, by 1946, in London. Strongly enamoured of the French, he would frequently socialise with French inmates of the German camps, dressing and eating like them and speaking their language. This behaviour was to pay off when, as the war was coming to an end, his German captors slaughtered all the Poles in his camp but releasing the French. Andrzejaczek escaped, portraying himself as French, and fled to Italy and, eventually, London. Interestingly, Andrzejaczek was to marry a Polish woman who was also a Francophile and a wartime underground agent.

Before their eventual meeting in Perth, WA this is geographically as close as Harris and Andrzejaczek got to each other. Both men were in London for the next handful of years, Harris completing a science degree and Andrzejaczek completing a diploma at the Polish School of Architecture. By 1956 Harris was Experimental Officer at the Greenwich Observatory and had married a civil servant in a neighbouring county parish church. By that same year Andrzejaczek had sailed to Australia under the Australian Government's Assisted Passage Migration Scheme - a government attempt to increase the country's population after the wars.

1963 finally saw the two men settled in Perth. Andrzejaczek had disembarked at Australia's largest city, Sydney on the east coast before finding temporary accommodations in Canberra and then moving to a short career in Adelaide. Opportunity and lifestyle brought him, and his wife Maria, to Perth where they lived till he passed away in 1987 (Andrzejaczek 2017). Harris accepted the post of Assistant to the PO Government Astronomer in 1957 and succeeded him in the job in 1963 (PO Arch). The next decade saw both men in the prime of their careers, before Harris passed away prematurely in 1974 (ADBHweb).

At this point, it is important to gain some global perspective. In 1964, the LO and PO had both been in existence for around 80 years and Australia's pre-eminent observatory at Mt Stromlo, near the eastern seaboard, for 40 years. The PO was having funding and light-pollution issues and was going through a tense period of negotiations with government about budgets and relocation. Several of the observatories that would comprise the IPPN were not even yet built, and existing observatories in South Africa and India that would eventually contribute to the IPPN, were reaching the end of their working lifetimes. The rebranded NASA was just a few years old and B&C, the telescope manufacturer, were looking forward to a 20-year anniversary. Architecturally, Post-Modernism was rejecting the austerity of the International Style and delivering buildings that reflected the local history, unafraid to be artistically expressive. A futuristic flair was becoming evident.

Two important, specific details arose from a 2017 interview with Andrzejaczek's son, Krysystof. First, the Australian National University's (ANU) online biography of Andrzejaczek describes him as "…*a dreamer and an idealist…*" who would not "…*compromise on deeply felt principles…*" (ADBTweb). This is not really true, according to his son. Rather, he was "…*very energetically driven to complete his projects…was very pragmatic…*" and "…*believed anything was possible…*" (Andrzejaczek 2017).

Andrzejaczek's ANU biographer, Charles Sierakowski, was a fellow graduate of the London Polish School of Architecture. Sierakowski's own 2012 biographical interview is online (Sierakowskiweb), and confirms that Andrzejaczek was Project Architect for the Perth Cultural Centre which is built in a post-International, Brutalist style (Apperley et al. 1994). But Krysystof was somewhat sceptical about the accuracy of Sierakowski's description of his father's motivations. Andrzejaczek's other well-known projects include the Kalgoorlie Court house, which is a Stripped Classical building and Perth's Anzac House, which has undergone many re-births (Apperley et al. 1994).

Second, according to Krysystof, Andrzejaczek was "…*absolutely fascinated…*" with La Grande Arche de la Défense that was being built in Paris during the last few years of his life. The architect for La Grande Arche, the Dane Johan Otto von Spreckelsen (1929-1987) used strong geometric shapes and would become Director of the Danish Royal Academy of the Arts (DKweb).

Andrzejaczek would have definitely known about Spreckelsen's work because, as Krysystof repeated, he was "…*very much in touch with what was going on around him in the world…*" often taking holidays to Europe and spending much time visiting and photographing buildings of interest, sometimes at the expense of including family (Andrzejaczek 2017).

The question of proportion was central to Andrzejaczek (Andrzejaczek 2017) and the history of astronomy, and in particular Johannes Kepler's tract in the 'Six-cornered Snowflake' (Kepler 1611) reveals an obsession with the Divine Proportion otherwise known as the Golden Rectangle. If a rectangle with length 'b' and width 'c' has:

$$\frac{a}{b} = \frac{b}{c} \cong 1.618 \qquad where \quad a = b + c$$

then it is said to be in Divine Proportions. This ratio is seen in very many places in biological morphology, and is thought to be a basic element of nature and therefore visually pleasing. It is seen in the proportioning of body parts in plants and animals and in such structures as the Parthenon and the Taj Mahal (Mathweb).

Coincidental as it may be, it is none-the-less instructive to consider that Andrzejaczek may have been unconsciously leaning towards these visually pleasing dimensions in the Perth-Lowell design. This may well be a simple human preference, but Andrzejaczek's personal history suggests he was particularly conscious of proportion. PWD records and PO archives show Harris' initial sketch, subsequent iterations, and Andrzejaczek's free-hand concept drawings all outside the Divine Proportion. Fig. 2 clearly shows this. If the final subjective decision to raise it to its present height had not been made, the structure would not have made Andrzejaczek or Harris, entirely happy.

According to Krysystof, Andrzejaczek "…*was a workaholic*…" but also a great socialiser, home entertainer, cook and raconteur. In Perth, he and his wife would often entertain political, social and academic luminaries, having dinner parties followed by long, intensive, boisterous, discussions, often centred around anti-fascism and culture. Andrzejaczek was also artistic, enjoying portrait painting (never buildings) and supporting compatriots like Perth Festival Director John Birman OBE (ADBBweb).

The ANU's biography web page on Harris was co-written by Muriel Utting who also wrote a history of the PO between 1896-1962 (Utting 2000). The other author of Harris' online biography entry was Prof. Philip Jennings, recently retired but still actively working on renewable energy and climate change at Murdoch University, WA (Jennweb). Unfortunately, neither Uttings history, nor Jennings personally, were able to add to the understanding of Andrzejaczek over that of the ANU biography but both agree Harris was active in Perth's local amateur stage-play scene (ADBHweb). Andrzejaczek's culture/art paradigm is therefore relevant to his relationship with Harris.

According to Haynes et al (1996) Harris could also be very…. pragmatic. On p93 they say he leveraged a government plan to build a freeway "…*straight through the observatory*…" by "…*threatening to delay this costly operation*…". The co-authors of that book are highly credible, hailing from the Commonwealth Scientific and Industrial Research Organisation (CSIRO), the Anglo-Australian Observatory (AAO) and the University of New South Wales. They would have been acutely aware of Harris's determination to retain functionality of the PO.

Large steps were taken between 1964 and 1968. Despite the fatal setback of AS-204 (otherwise known as Apollo 1), NASA was hitting its straps with the Apollo Program and about to fly around the Moon. They were also setting their sights on Mars with the Viking Program (Nature 1969) and would soon engage the LO to develop the IPPN.

Harris was simultaneously negotiating with the Bergdorf-Hamburg Observatory, in Germany, to build a meridian transit dome for a visiting survey team (PO Arch, Bowers

2017), as well as expedite the recently approved move of the entire PO its new site. At 380 m altitude, the new location in the Darling Ranges is seven times higher than the previous site (ElevMapweb), light pollution from the city's half million residents (UDIAWAweb) being the primary reason for the move (Bowers 2016). Though Harris had no inkling of the opportunity coming his way over the next few years, he would have been pleased to have a new, raised lot away from the general population.

Andrzejaczek was involved in the design of the meridian building on the new PO site, and although how exactly he became involved remains unknown, there is documentary evidence that he was highly regarded and known by Harris. In a 1965 letter from an observatory manager in Hamburg, Dr J von der Heide, to Harris, Heide remarks that Andrzejaczek's design of the new Meridian Dome was "…*as technical as artistic very good building. We are very indebted to you and we beg to say our high respect to Mr Andrzejaczek.*" (PO archives cited in Bowers 2017).

Around this time, new telescope facilities were being built at Cerro Tololo, Chile (1966), Mauna Kea, Hawaii (1967) and Kavalur, India (1968). These would become part of the IPPN, the Coordinator for which was William Baum. Harris was dedicated and active, and made interstate lobbying trips on behalf of the PO, so he was well known to the Director of the Mt Stromlo Observatory, Dr Bartholomeus J. Bok. When Baum visited Mt Stromlo, Bok recommended he approach Harris to discuss inclusion of the PO in the IPPN. The subsequent introductory letter from Baum to Harris is dated 1968 July 19 (PO Arch).

Those were heady days. Frank Borman, James Lovell and William Anders were preparing to blast off to the Moon, Harris was in his new office, successfully enticing international astronomy programs to the PO, and Andrzejaczek was a highly regarded local architect with experience in designing telescope facilities. Exactly 1,000 days later the Perth-Lowell telescope saw its first stars. But for now, the pressure was on.

Baum's letter explained that the best telescope for the job was a 24-inch B&C Cassegrain reflector with focal ratio F/75. LO would build a 35 mm film camera to take 4.5" images, in quick succession, through a range of inline filters (PO Arch).

It was more than 12 months of to-and-fro communications between Harris and the LO, and between Harris and Treasury, before in September 1969, the business manager of the LO suggested the loan of the B&C telescope for use in the IPPN, and that it may be retained for subsequent use. The official offer came late in 1969 and included funds for two local operators, equipment, supplies and petty cash. The telescope would be fabricated in five months for a June, 1970, delivery (PO Arch, LO Arch). But the PO was responsible for building the housing for the telescope.

There are four basic parts to a telescope installation. A solid stand (or pillar) on which sits the particular telescope mount, attached to which is the telescope or Optical Tube Assembly (OTA). All of this is housed in an appropriate structure, the design of which varies hugely but addresses such factors as the science, budget, protection from weather, optimizing functionality and, perhaps, aesthetics. But in the late 1960s, who knew the logistics of such an undertaking? And how would a scientist in far-flung Perth, WA, find the latest and best? Of course, the US Department of Defence was only now conceiving of the Advanced Research Projects Agency Network (ARPANET) that would become the global internet of today. This was an era of paper-based mail, plane trips and conferences, especially so for scientists like Harris who suffered the tyranny of distance.

In a September, 1969, letter from Harris to the LO, he asked for details to allow him to design the housing. He asked for specific housing requirements of the telescope including pillar height, axis location, OTA length and minimum internal radius of the dome. In this no-nonsense letter he warned the LO that results would be seriously impaired by ground-level conditions. This is the first inkling that he wanted to have a tall structure. To keep costs down and to reduce wind loads, he suggested an octagonal dome with hinged flap shutters. He was also very straight forward about his views on the bureaucratic and slow-moving work force in Perth's PWD (PO Arch).

Harris wrote to his Treasury Department outlining the specifications of the telescope, and 2 weeks later he was writing his acceptance letter back to the LO. His discussions with Treasury were far from over, but, in the acceptance letter to the LO, Harris expressed gratitude for the generous offer and described his approach to the project. He was already preparing sketches and would forward them to the PWD, as soon as possible, for them to do their own drawings and costings. He again warned that the PWD plans may take 12 months just to get approval, and that their workers were very slow. He made it clear that he was still engaged in bureaucratic discussions with Treasury and also that parliament would have to approve the expenses (PO Arch).

Harris subsequently learned that NASA's telescope supplier would be B&C in conjunction with the LO, and that the OTA required a dome with internal radius 11 ft 6 inch (3.5 m). He set about designing the facility. His original sketch of the Perth-Lowell tower is a simple pencil drawing on notepad paper (second from left in Fig. 2). The design is reminiscent of the building in which he had spent much of the previous few years - the Government Astronomer's observatory/residence at the old facility near Perth city. He drew plans for a rectangular building 28 ft x 23 ft with a second storey floor 24 ft above the ground. The access stairwell, within the building, was offset to one side. Curiously, the 'dome' was an octagonal structure with internal radius just 8 ft 6 inch. It was reminiscent of the old PO's 4-sided tower cupola.

In an undated letter of around this time, Harris lays out the requirements for the installation. The letter is likely intended for the PWD because it was accompanied by his sketches and a promotional brochure from B&C. In it he makes a few stipulations and a few concessions. The structure was to be built on the "…*central ridge of the Bickley site…*" some 300 ft north of an existing astrographic telescope dome.

Harris planted a confusing seed for the potential to increase the tower height by saying that, on economic grounds, the ideal height of the telescope axis had been reduced from 40 ft to 31 ft. No documentation was found to explain the increase from 24 ft to 40 ft, then back to 31 ft. The building would now have two storeys of 12 ft, plus 7 ft for OTA and mount. Critically for the design outcome, Harris advised that his sketches are for a rectangular building "…*on the assumption that the added cost of corner bricks is less than that of skilled labour for the curved end-walls of existing telescope buildings.*" Curved walls were never mentioned again.

In this initial proposal, Harris made another big concession to the budget. His obvious preference for a hemispherical dome, that would minimise wind resistance, was mentioned but his octagonal design was offered to reduce construction costs. He also explained his first approximation for the design of service rooms and utilities. Ever the pragmatist, he describes two more detailed aspects of the building. A 5 ft square hatch in the floor of the top level was to accommodate raising/lowering telescope parts and other equipment via a block and tackle hoist mounted inside the dome. He planned the arrangement of internal doors with respect to bearing the load of the dome down through the structure to the ground. Harris was clearly more than an academic. He was also a good construction manager.

That letter (PO Arch) was likely dated late 1969 because Harris then wrote to Baum in late January, 1970, outlining exactly the same details. He also advised that he had met with Andrzejaczek just after Christmas 1969, and was waiting for a quote from the PWD. Presumably, that meeting with Andrzejaczek was the crucial point when both men had an understanding about the 'best' height for the facility.

The preliminary cost estimate of $62,000 arrived February 11, 1970, and was based on first-round drawings done by Senior PWD Draftsman Mr A. Chinnery – selected by Andrzejaczek for his experience in drawing plans for the main buildings of the new PO (PWD Arch). Within a few days, Harris wrote to Treasury asking for funding, and although he was scratching around for ideas to use the telescope after the IPPN program, he was advising Treasury that it was unlikely the telescope would need to be returned to NASA.

April 1970 was likely Harris' most stressful month. In a stream of carefully worded letters swapped between himself, several Treasury financial officers and the Under Secretary of the WA Government's Finance Department, the value of spending the equivalent of a total annual PO budget on one installation, drew some revealing and depressing sentiments from the bureaucrats:

- *"…observatory work is mainly of scientific interest and the principal benefit to the State is one of prestige…"* (Finance Officer P. Wilson);
- *"…there would be no direct benefit to the State…"* (Wilson);
- however, outlaying $62,000 for a $90,000 facility was appealing; and
- *"…the advantages are intangible and support is not given…"* (Chief Finance Officer K. Binks).

Despite the politics, by May 20, 1970, Harris received approval from Treasury, acceptance from Baum, approval from NASA and had signed a Memorandum of Understanding. He had specifications for the camera that the LO workshop was building, specifications for the telescope and dome, and knowledge that B&C were in the design and fabrication phase. He had also reminded Andrzejaczek of the first light deadline of April 1, 1971.

Everything looked good, but Harris' first sketch had no supporting scientific documentation to describe the technical pros and cons of height, thermal ground effects, wind turbulence, orientation, humidity. Was any of this considered?

A CalTech alumnus, and Chief Engineer at B&C (B&Cweb), Mr William Baustian worked for Lick and Kitt Peak Observatories, and was probably one of the best-known experts on the logistics of building large telescope facilities. Baustian was very experienced, and wrote widely about many aspects of setting up all aspects of a research observatory (see Baustian 1966, Baustian 1968, Baustian 1970), but he did regard the telescope housing as "…*a necessary evil*…" (Baustian 1971). So, although there was knowledge in the international community, no evidence was found that it was sought by Harris or Andrzejaczek, or even suggested by NASA or the LO. It is interesting that Baustian was very well known to B&C and the LO and presumably to Baum, but no evidence was found that Harris communicated with him about specifications for the Perth facility.

B&C are no longer in business but do have a web site built by ex-employees and enthusiasts. One of their web pages (see B&Cweb) has links to installations of their telescopes around the world. Photographs of telescope housings are not prominent in these pages but they do have a good representation of the Siding Spring Observatory (SSO) in New South Wales, Australia, which in the 1960s also housed one of B&C's 24-inch reflectors (B&C1web). A quirk of these telescopes, is that they require an off-set pier, hanging, in the southern hemisphere, off the south side. This was confirmed by Peter Verwayen, Senior Operations Officer at the Research School of Astronomy and Astrophysics, SSO (Verwayen 2017) and adds to the required design considerations around the flexure of the exposed concrete pier for the Perth-Lowell. After installation, vibration issues did plague the performance of the telescope but they were not due to the engineering of the pier (PWD Arch). The metal wheels that rotated the dome caused too much vibration and were eventually replaced with pneumatic wheels (PO Arch 2017). But all this was ahead of Harris and well after first light.

There is no evidence that Harris spoke about installations for the new telescope to Bok at Mt Stromlo. This is surprising considering Bok had written an article for the Journal of the Royal Astronomical Society of Canada in 1960, in which he identifies at least four other locations in WA that had significantly more clear nights per year than Perth (Bok 1960). A draught of this journal paper is held in the LO archives as part of the Perth-Lowell project records. The other sites are in remote rural towns north and east of Perth where the climate is hot and dry, but clearly they were not seriously entertained as sites for NASA's IPPN installation.

Astronomers of the day had one stipulation: the new dome should be "*high enough to avoid ground-level seeing problems*" (PO Arch). In a letter from Harris to the LO, he hints that the new dome should be higher than the existing University telescope building but there is no evidence that the seeing from this telescope was a particular problem.

Features that make the Perth-Lowell structure unique are; physical isolation from other buildings, the external stairwell, the curious height and the exposed pier. Could there be similar examples elsewhere in the world that either Harris or Andrzejaczek might have seen? As Andrew Williams, retired astronomer from the PO, said in an unpublished Statement of Significance: "*Elevated domes for telescopes are extremely rare, and only a handful exist worldwide, making the Lowell dome a truly unique building.*" (Williams 2016).

World-wide, there are less than a dozen observatories with a tallish tower that isn't the smaller part of an incorporating building. Most observatories are built to the hemisphere-atop-a-cylinder design, with an internal stairwell to a second storey observing platform. There are some unique designs like the expressionist Einstein Tower in Potsdam, designed in 1921 by German Erich Mendelsohn (Mendelsohnweb). The Tuorla Observatory in Finland is tall and isolated, but again, a dome atop a cylinder (Tuorlaweb). And the Marina Towers Observatory in Wales is tall, on a cubic building, and isolated - but the building encloses the stairs (Marinaweb). The extraordinary Sphinx Observatory on the Jungfrau peaks in Switzerland has some common attributes, but again has internal stairways (Jungfrauweb).

There are a few basic guidelines for telescope facility construction, but local conditions dictate engineering specifications. The Director at Armargh Observatory (Ireland) and the Project Manager at Stellarium Gornergrat (Switzerland) agree that geographical location is the primary consideration, followed by local placement and general seeing optimisation (Burton 2017, Riesen 2017). It seems Harris was doing in the 1960s exactly what any good astronomer would have done.

Site location is obviously important. Peter Birch, PO astronomer during the IPPN program, believes the tower was raised to such a height in part to be able to see over the existing forest canopy and track Mars as far as possible towards the horizon (Birch 2017). This is partly linked to the question of why the dome was situated exactly at a height contiguous with the tree canopy - a height at which lamina air flow is unlikely, qv turbulence effects on wind turbines (see Sagrillo 2009). Sagrillo (2009) offers the common and generalised rule that the whole of a wind turbine must be clear at least 10 m above any structure that is within 150 m. The 40 ft (12 m) high dome floor of the Perth-Lowell is 10 m away from 15 m tall trees on at least two sides. It is in turbulent air amongst the forest canopy!

In any case, to prepare the site for construction, bulldozers arrived the first week in July, 1970 (PO Arch), the Eucalypts of the Jarrahdale State Forest presenting no real obstacle, even providing marketable timber (PO Arch). The new site was higher, darker and more remote than its predecessor, and clearing land for the new facility simply required Harris' request from the lessor of the PO property, the WA Forests Department (PO Arch).

As the clearing began, two contemporaneous letters give perspective to the story. First, Andrzejaczek's full report is dated July 10, 1970 and constitutes the most complete description found in the records. In it, he describes several criteria best paraphrased here:

- the pier foundation will descend 10 ft into the ground and sit on "*ironite rock subsoil*";
- no explosives to be used;
- a hoist with 2-ton capacity will be installed inside the dome above a 5 ft square hatch;
- to reduce thermal turbulence from roofing, ground level buildings on the south side will be excluded from the design;
- the dome floor will now be 40 ft 3 inches above ground;
- to support the dome floor, two vertical concrete walls will rise alongside the otherwise exposed telescope pier that will be 2 ft 6 inches square;
- horizontal, concrete 'fins' will connect the two 'legs' and act as solar

- irradiance shields;
- extra concrete foundations will be laid in the event that walls of perforated bricks are subsequently required to protect the pier from strong winds;
- a 1-inch gap between the pier and all parts of the structure, including the dome floor, will result in "…*no transmission of vibrations from the supporting structure onto the pier.*";
- the dome enclosure would be a low thermal capacity metal/plywood structure, the cavity of which would be ventilated;
- the dome itself would be a double skin of sheet metal; and
- amenities building with steel roof located at ground-level on east side (PWD Arch).

Clearly, he and Harris had been in close conversation about what was actually going to happen. The height was back up to 40 ft 3 inches and Andrzejaczek had designed a unique concrete structure to support the dome floor. Also obvious, is that wind turbulence and thermal seeing was being seriously considered.

The second letter was from Harris to Baum and dated July 13, 1970. Harris advised that the "…*structure has grown…*" partly because the architect thought the appearance would be more pleasing if it is taller. Herein lies the crux of the present investigation. Around June, 1970, Harris and Andrzejaczek were zeroing in on their preferred design and Harris was now boldly presenting it to NASA's representative. He was simultaneously telling Baum that marketable trees were removed from site, clearing was underway, and (still hedging his bets) there'd been a misunderstanding by Andrzejaczek when he'd read a letter from the LO Business Manager in which was made a statement about "…*favourable times for planetary observations…*".

Harris' April 1, 1970, first light target was 260 days away and no foundation holes had yet been dug. Andrzejaczek was still hoping for a completion date of March 1, 1971, but his department's first estimate was for a 34-week project starting August 3, 1970 and ending March 31, 1971. More modifications were coming.

In August, there were significant onsite changes and the revised quote had grown to double the original. An insulated, metal, hemispherical dome was to be built by the State Engineering Works and there were changes to the hatch and many aspects of the ancillary mechanisms. The height was now going to be 37 ft 4 inches, though this conflicts with Andrzejaczek's month-old report. In the end, a $12,500 dome was to be delivered to site and $43,000 diverted from a Government gaol-extension project in the northwest city of Broome (PWD Arch). A 2017 personal measurement of the tower by the PO Outreach Manager, Matthew Woods, concurs with the 40 ft 3-inch height to the dome floor.

Vegetation was removed and pruned to height around the new site. Plans obtained from the WA Government under the Freedom of Information Act 1992 (Fig. 2), show that all trees were removed from a rectangle with its western side 60 ft from the dome, southern side 35 ft from it, eastern side 35 ft and northern side 40 ft from the dome. To the south is a large open area where trees over 30 ft were removed. Outside the exclusion zone on the north side, trees over 50 ft were removed. See Fig. 3.

The WA Government's State Records Office (SRO) archives hold engineering

documents describing wind loads on the concrete pier and design features to mitigate wind stress and solar thermal stress. Shortly after completion, annoying vibrations of 5" were considered too much, and efforts made to reduce it to 2". The physical structure of the Perth-Lowell was built to rigidly withstand movement by the wind, for instance designing the wrap-around external stairs and the fact that 50% of the budget went to concrete, half of which comprised the 26 $m_3$ telescope pier (PWD Arch). At one point, attempts were made to stabilize the wind vibrations by placing steel props around the outside of the structure (PWD Arch).

Placing the dome at canopy height was, perhaps, inadvisable. Although perforated concrete blocks were proposed to ameliorate wind loads on the pier, thermal turbulence from the ground was considered the significant, constant impactor on seeing. Contemporary materials and technology can produce a reasonable telescope dome on a tower with an exposed pier (Kattner 2017) that also allows the local winds to cool ground thermals (see Orionweb). In the 1960s this was likely possible but would have been prohibitively expensive.

No evidence was found to indicate pre-construction optical seeing measurements being taken. In-fact, in two letters from LO to Harris, a laissez faire attitude was evident. On September 22, 1969 the LO Business Manager said simply "…*some type of shelter would be required*…", and on December 5, 1969 the same person said they were not concerned about the type of housing, just that it be "… *high enough to avoid ground-level seeing problems. If strong winds are common, perhaps some sort of windscreen could be provided*…" (LO Arch). Harris' response was a little more specific, stating in his same September, 1969, letter mentioned above, that the floor level of the new building should be "…<u>at least</u> as high as the existing telescopes…" the words "*at least*" being underlined. This presented the possibility that seeing from the existing 16-inch telescope may have been compromised, but log books and archives indicate no such issues (Bowers 2017). See Fig. 4.

Former PO astronomer Andrew Williams has no knowledge of any seeing measurements being taken before construction (Andrews 2017). The first evidence for such records appears in the Perth-Lowell telescope Log Book entry of April 10-13, 1971. That log has an entry indicating the first night's seeing was too poor to observe, and the following few nights were clouded out. The first entry of internal dome conditions appears early evening of April 16, 1971. Dome temperature, humidity, transparency (0-10) and scintillation (0-10) were logged that night (PO Arch).

The European Southern Observatory provides a good and credible coverage of the basics of atmospheric seeing (Zagoweb), and lists three main contributors:

- scintillation - mainly caused by air movements at around 12 km;
- turbulence in an interface layer between 30 m and 500 m; and
- ground and structure turbulence up to 30 m.

Starting with these bulk atmospheric qualities, a seeing index may be constructed as a Full Width Half Maximum of an averaged disk of light taken during a 10-30 s exposure. For modern telescopes, a FWHM seeing value below 1.0" is good and 2.0" is acceptable. The telescope housing itself can contribute as much as 2" (Zagoweb). In 1972, the Perth-Lowell was still trying to get the FWHM under 5" and aiming for 2"

(PWD Arch).

Other measures of seeing like the Fried Parameter (Fried 1966) and Strehl Ration can be calculated, but these tools were only just appearing in the literature in the very late 1960s (see for example Hazra 1975). Zago pointed out in 1995 that there was no way to directly measure the factors that impacted on seeing immediately around the telescope. Scintillation seeing at the Perth-Lowell was assessed objectively on a 1-10 scale (Birch 2017), similar in concept to the Pickering Scale (AstroMagweb) but without recourse to actual measurements. Transparency was also subjectively assessed on a 1-10 scale as a measure of "…*how dark the sky appeared to be*…" (Birch 2017).

Construction had been set to start on August 3, 1970 and be completed, including the dome by March 31, 1971. This clearly did not allow time for testing before the first Mars observation run on April 1, 1971. The 40 ft height is also mentioned in Harris' letter to Treasury on February 16, 1970 and then by July 13, 1970, a week after the trees had been bulldozed prior to construction, Harris advised Director Baum at the LO that Andrzejaczek "*unfortunately*" had misinterpreted a letter from the LO Business Manager to mean there was plenty of time for the construction project (LO Arch, PO Arch).

From around this time, any documented evidence about the construction essentially stops. The next communications are about post-construction issues around vibration mitigation and the logistics of setup and operations.

On August 29, 1970, the WA public were made aware of the project in a newspaper article that, presumably after an interview with Harris, informed of the need to withstand gale-force winds to avoid ground thermals, to shade the pier and to have a low thermal capacity dome. The article says that as of that date, excavation of the foundations had been started. Fig. 5 shows the newspaper article alongside a picture of the almost-finished product. The latter does show some distance between the dome and the treetops, but clearly indicates that canopy-level air flow would impinge on the dome.

Around this time, there was plenty of communication between Harris and the LO about operational procedures, and an inventory of items to be shipped was sent in October, but virtually no documentary evidence of the actual construction process. By years end 1970, B&C were packing up the telescope for sea freight. The LO paid B&C just under US$ 9,000 to transport the telescope and have a B&C technician on site for the installation. An early estimate of the telescope's arrival was November, 1970, but by February 22, 1971, shipping agent Loretz & Co., advised the telescope was sent to dock for loading on January 22, and left dock around January 25 onboard the SS Gertrude Baake (LO Arch). It was six weeks before it arrived at Perth's main harbour, Fremantle, then another 2 weeks in customs and processing.

After clearing customs, the telescope was eventually onsite in late March. It was installed around April 5, 1971 and the aperture opened on the evening of April 8, 1971. Documentary evidence to elaborate on these few weeks was not found, but inclement weather delayed observations until at least April 10, 1971 (PO Arch).

**CONCLUSION**

The Perth-Lowell telescope installation was a late 1960s idea originating from NASA's need to have 24-hour, global observations of Solar System planets including Mars. Particularly needed was continuous atmospheric observations in anticipation of the upcoming Viking missions. A network was required, and locating one in the south west of Australia would be beneficial, especially if it could be arranged before favourable planetary alignments in 1971. Several sites in the southern half of Western Australia had been appraised in 1960 by the Director of the premium observatory on Australia's east coast – Dr Bartholomeus Bok from Mt Stromlo.

The Lowell Observatory in the USA was to operate the IPPN for NASA, and through astronomy conferences and international connections, Bok had recommended the Perth Observatory. Over a 3 year period of professional, persuasive and pragmatic negotiations, the Perth Observatory's Bertrand Harris managed the construction of a tall, isolated, post-modern telescope housing that saw first light 1,000 days after he was first contacted by the Lowell Observatory. He appears to have completed this task in close association with one of the WA Government's quality architects, Mr Tadeusz Andrzejaczek.

Unknown to each other, Harris and Andrzejaczek both studied and worked in London for a handful of years either side of 1950, but they didn't meet until they were both settled in Perth; Harris by 1957 and Andrzejaczek by 1963. By then, each had lived very different lives, but each brought similar qualities to the Perth Lowell project.

Andrzejaczek was born into poverty in war-torn Poland, raised by grandparents and injured within a week of enlisting for WWII. A survivor of six years in German concentration camps, he made his way to London where he qualified as an excellent architect. He married a similarly strong and politically aware woman. He was inspired strongly by French culture and closely followed architectural styles of the day

Qualifying for sponsored migration, he and his family boarded a ship for Australia. He had a short career in South Australia then moved to Perth where he designed many public buildings including court houses, office blocks, police stations and art centres. Andrzejaczek was an avid portrait painter and socialised frequently with politicians and leaders in their fields. His son described him as pragmatic but having the belief that anything was possible. He was very aware of world politics, culture and especially design.

Growing up in 1930s south east England, Harris had a less traumatic youth, was a high achiever in a private school and married in a nearby parish church. He joined the Royal Navy and began his career with the Greenwich Observatory in London and soon accepted an offer as Assistant Astronomer in a small, remote city on the other side of the planet.

Harris proved his mettle in negotiations with authorities and energetically expanded the Perth Observatory's international programs at the same time as overseeing the relocation of the entire observatory to a new site. Battling disinterested politicians and severe budget threats, he nonetheless employed Andrzejaczek to design several of the new telescope housings, including the object of this research. With glowing compliments from internationals about Andrzejaczek's work, it seems that Harris had already cultivated the ideal person for the Perth-Lowell telescope job.

When the NASA-funded IPPN idea was presented to him, Harris would have immediately seen the benefit not only to the Perth Observatory, but to WA's future as an important astronomy centre. NASA was offering a big, expensive telescope on indefinite loan, and all he had to do was build a housing for it. He had a new site ready to go, but he also had a short timeline.

Before the internet and before easy access to peer reviewed knowledge about such projects, Harris and Andrzejaczek relied on their personal drive, experience and strength to manage a project that was like no other in the world. These men had vastly different life experiences, but both were technical and artistic, professional and strong. Harris had his hands full with politics and international negotiations, while Andrzejaczek had the deserved respect for his designing skill and for having his 'finger on the pulse' of contemporary culture. Both had high self-confidence, a no-nonsense approach, and a regard for each other.

Outside of Harris and Andrzejaczek, it is not evident that expert knowledge was sought or offered for this project. Ground thermals, wind turbulence and mechanical vibration issues were all addressed with subjective confidence, the project apparently growing as and when the opportunity was seen by either of these two men. It's true that Harris later lamented the fact that he "…*did not take sufficient care in the planning stage to ensure that the pier was rigid enough*…" and wanted to "…*discuss this … with people who have experience in the design of telescope dome buildings and the associated stability requirements*." (POArch). But this is 20/20 hindsight. He was very time-poor during the construction management phase, and international research, as slow as it was in the late 1960s, would have prevented the deadline being met.

Although it is also true that Harris quoted Andrzejaczek as having wanted the installation to be taller so it was more 'pleasing' to look at, the evidence suggests that in their regard for each other's technical, cultural and social idealism, it was realised that Harris' purposes were served best with the highest possible dome, and Andrzejaczek's input resulted in Divine Proportions (knowingly or not). Harris' budgeting skills assured the funds to achieve the best outcome.

In 1,000 days, the Perth-Lowell design went from a 24 ft tall cubic structure with an octagonal dome and hinged fabric shutter, to a 40 ft 3 inch Post-Modern concrete edifice with an automated, hemispherical dome containing a 2-ton hoist and a brand new 24-inch Boller and Chivens reflector. It had its post-installation problems, but, as of 2020, is still onsite under a current Memorandum of Understanding with NASA. It continues to be a highlight of visitor tours for schools, the general public and on both dark-sky and clouded evenings.

**FIGURES**

| Andrzejaczek | Year | Harris |
|---:|:---:|:---|
| Born southern Poland | 1915 | |
| Moves to Bydgoszcz with grandparents | 1916 | |
| | 1917 | |
| WW I ends | 1918 | |
| Poland re-created by Versailles Treaty | 1919 | |
| | 1920 | |
| | 1921 | |
| | 1922 | |
| | 1923 | |
| | 1924 | Mt Stromlo Observatory opens in Australia |
| International architecture style arises in Europe | 1925 | Born southern England |
| | 1926 | |
| | 1927 | |
| | 1928 | |
| | 1929 | |
| | 1930 | |
| | 1931 | |
| | 1932 | |
| Moves to Warsaw | 1933 | |
| | 1934 | |
| | 1935 | |
| | 1936 | |
| | 1937 | |
| | 1938 | |
| German massacre in Bydgoszcz | 1939 | WW II begins |
| Enlists and leg shrapnel then German holding camps | 1940 | |
| | 1941 | Matriculates Royal Grammar School, Guildford |
| | 1942 | |
| | 1943 | Joins Royal Navy |
| | 1944 | |
| Out of holding camps | 1945 | WW II ends |
| Living in London | 1946 | Employed at Royal Greenwich Observatory |
| | 1947 | Boller & Chivens founded |
| | 1948 | |
| | 1949 | |
| | 1950 | |
| Graduates Polish School of Architecture, London | 1951 | |
| | 1952 | Graduates B. Sc. Uni London |
| | 1953 | |
| Arrives Sydney Australia (Perth stopover) | 1954 | |
| Sydney to Canberra to Adelaide | 1955 | |
| | 1956 | Promoted to Experiment Officer Greenwich Obs. |
| | 1957 | Assistant Astronomer at PO |
| | 1958 | NACA merged into NASA |
| | 1959 | |
| Bok's article on potential Australian observatory sites | 1960 | Apollo era begins |
| | 1961 | Acting Govt Astronomer at PO |
| | 1962 | |
| Arrives Perth to stay | 1963 | Becomes Govt Astronomer at PO |
| | 1964 | |
| Highly praised for Meridian dome design | 1965 | |
| Cerro Tololo Observatory opens | 1966 | PO moves to Bickley |
| Mauna Kea Observatory opens | 1967 | Harris gets Meridian program running |
| Kavalur Observatory opens | 1968 | Baum writes to Harris |
| First Moon landing | 1969 | LO confirm shelter needed |
| Full architect report July | 1970 | Bulldozers arrive July |
| Telescope arrives Fremantle Mar | 1971 | Telescope installed and first light April 8 |

*Fig. 1: Timeline showing how Andrzejaczek's and Harris's lives merged.*

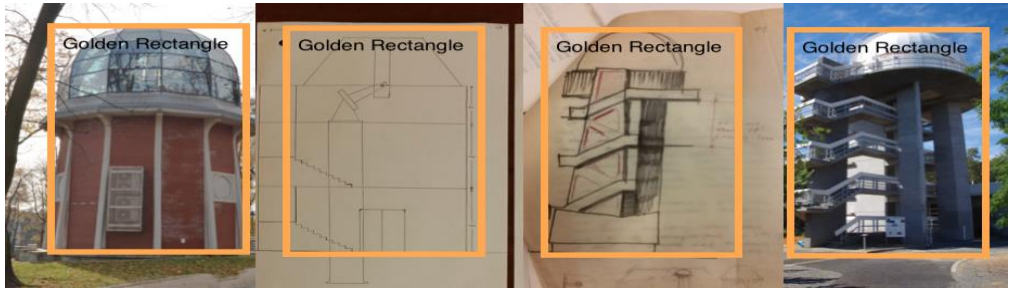

*Fig. 2 The Divine, or Golden Rectangle, in orange, is a visually pleasing shape seen in nature and used in architecture. The observatory in Częstochowa (far left), conforms to this shape, as does the final Perth-Lowell facility (far right). Harris'*

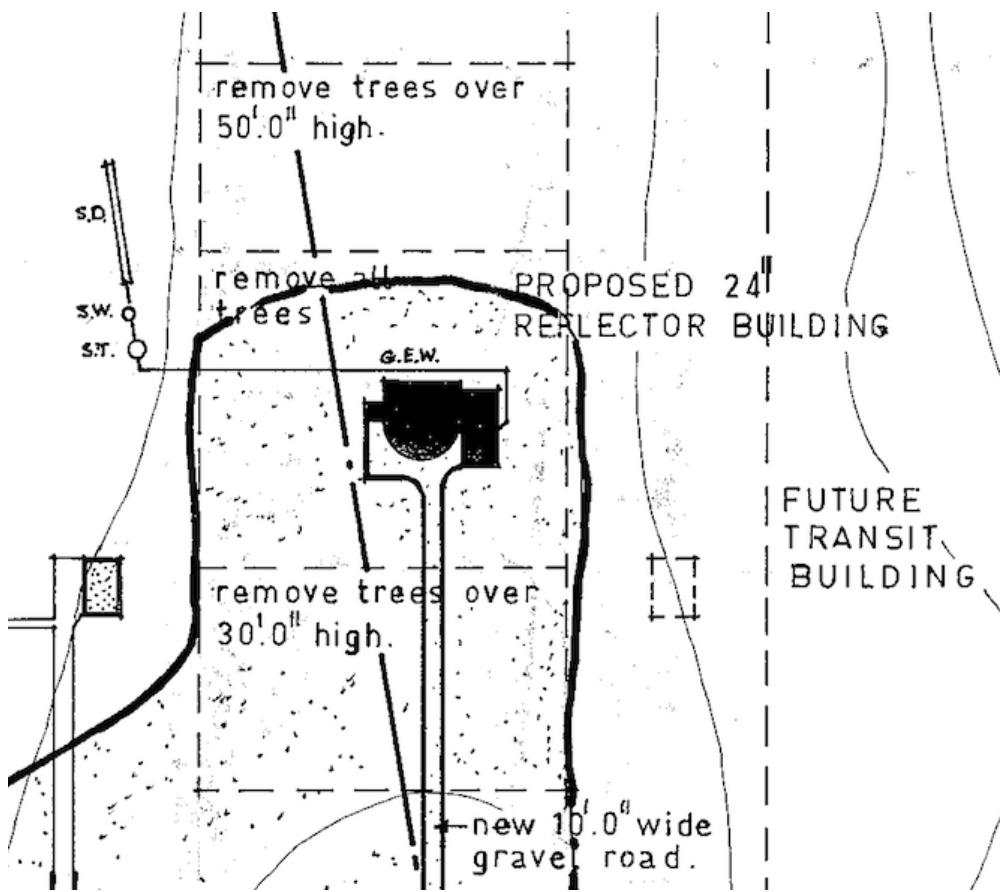

*Fig. 3 Part of the site plan for the Perth-Lowell facility which is shown in solid black. All trees and vegetation were removed from a rectangle immediately around the tower, particularly up to 60 ft to the west. Trees over 50 ft were removed outside a line 40 ft to the north, and 35 ft to the south. Today, trees of at*

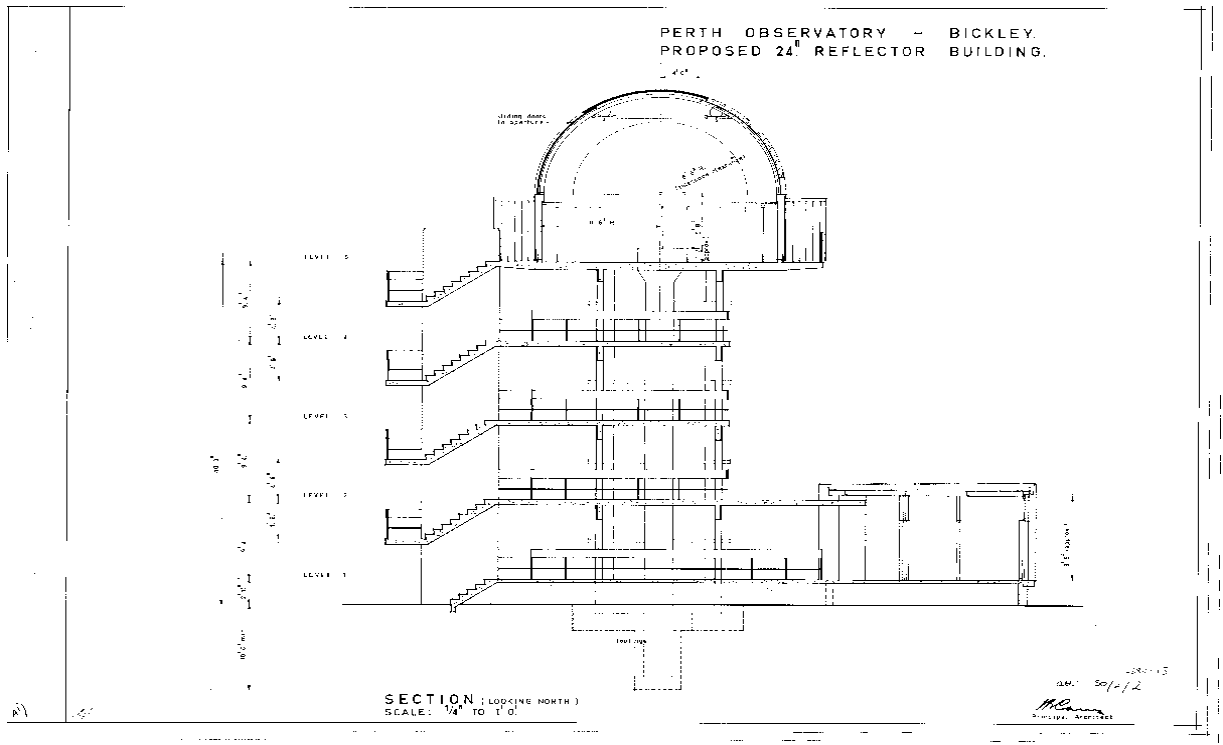

*Fig. 4 Final plans for the Perth-Lowell telescope facility. The height of 40 ft 3 inches is clearly shown, as is the exposed pier and wrap-around staircase. Shade fins can be seen as horizontal members behind the pier. The supporting concrete legs are not shown in this cross-section. (Credit: Cann, PWD)*

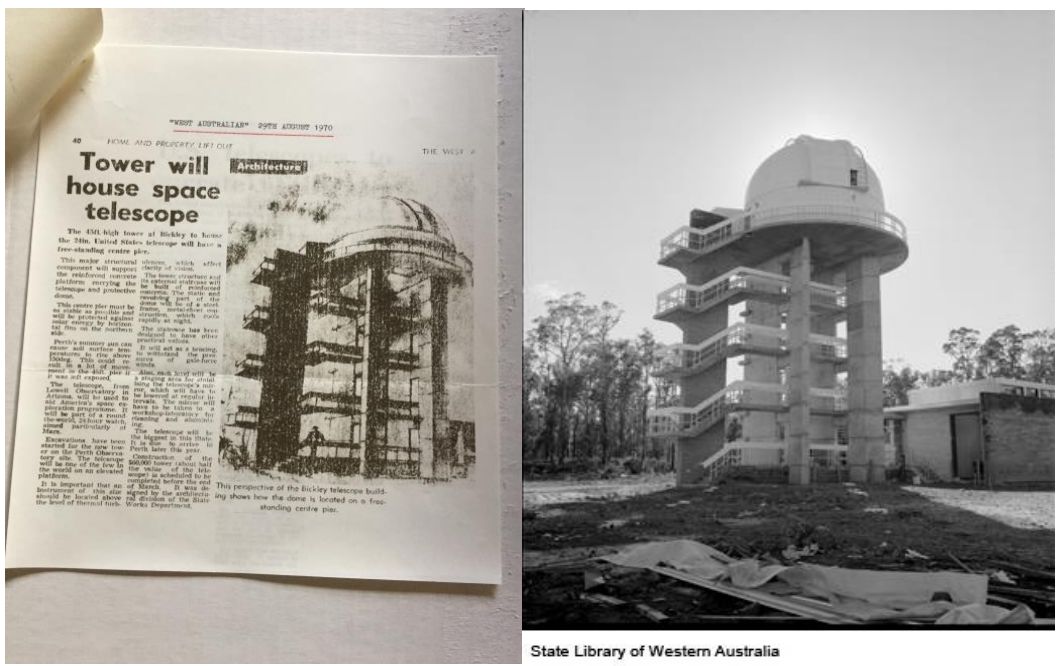

*Fig. 5 Left: On 29 Aug. 1970, the WA public was promised a new Perth-Lowell telescope installation. Other than some journalistic errors of fact, the article uses one of Andrzejaczek's later drawings that captures the final outcome very closely. Right: The facility was open for business 230 days later.*